\documentclass[prl, 
twocolumn,
showpacs ,amsmath ,aps]{revtex4}
\usepackage{times}
\usepackage{natbib}
\usepackage{graphicx}
\usepackage{bm}
\begin{document}
\title{Nano Positioning of Single Atoms in a Micro Cavity}
\author{Stefan Nu{\ss}mann}
\author{Markus Hijlkema}
\author{Bernhard Weber}
\author{Felix Rohde}
\altaffiliation[Current address: ]{ ICFO--Institut de Ci\`{e}nces Fot\`{o}niques, Jordi Girona 29, Nexus II, 08034 Barcelona, Spain}
\author{Gerhard Rempe}
\author{Axel Kuhn}
\affiliation{Max-Planck-Institut f\"ur Quantenoptik, Hans-Kopfermann-Str. 1,  D-85748 Garching, Germany.}
\date{\today}
\begin{abstract}
The coupling of individual atoms to a high-finesse optical cavity is precisely controlled and adjusted  using a standing-wave dipole-force trap, a challenge for strong atom-cavity coupling. Ultracold Rubidium atoms are first loaded into potential minima of the dipole trap in the center of the cavity. Then we use the trap as a conveyor belt that we set into motion perpendicular to the cavity axis. This allows us to repetitively move atoms out of and back into the cavity mode with a repositioning precision of 135\,nm. This makes possible to either selectively address one atom of a string of atoms by the cavity, or to simultaneously couple two precisely separated atoms to a higher mode of the cavity.
\end{abstract}


\maketitle

Worldwide, intense research is devoted to control the position of an array of single atoms. To meet this objective, several techniques are presently explored with optical \cite{Kuhr01,Dumke02, Mandel03,Bergamini04,Sauer04}, magnetic \cite{Haensel01:2, Folman02, Horak03} or electric \cite{Guthoerlein01,Mundt02,Kielpinski02} fields. However, none of the experiments performed so far has achieved positioning in combination with strong coupling to a high-finesse optical cavity. Either the atom could not be stopped inside the cavity, or strong coupling was not achieved, or trapping of a discrete number of atoms was probabilistic and cavity addressing was not possible \cite{McKeever03, Maunz04,Boca04,Maunz05}. The problems encountered when combining single-atom control with strong coupling are twofold: On the one hand, the small optical cavity required to achieve strong coupling adds a significant complexity, making the experiment challenging. On the other hand, atomic motion in the regime of strong coupling is dramatically different from free-space motion \cite{Maunz04,Munstermann99:2}. We now report on an experiment where the novel phenomena that a strongly coupled atom experiences inside a cavity are exploited to achieve the above-mentioned objective. Our experiment is the first in which a distinct number of atoms is repeatedly transported in and out of and brought to a halt inside an environment where atoms and cavity can no longer be considered separate entities by virtue of their strong coupling.

 
\begin{figure}
\centering\includegraphics[width=0.9\columnwidth]{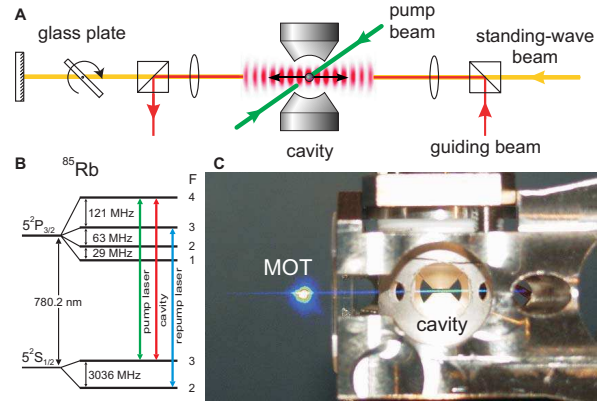} 
\caption{\textbf{Experimental setup:} ({\bf A}) Sketch of the dipole-trap arrangement that is used to guide atoms into the cavity. A standing-wave dipole-force trap allows us to freely adjust the position of an atom within the cavity mode by tilting a thick glass plate in front of the retro-reflecting mirror. Atoms that are trapped in the antinodes are displaced accordingly. (\textbf{B}) Level-scheme of $^{85}$Rb including the relevant transitions. (\textbf{C}) Side view of the cavity (cone-shaped mirrors), superimposed with absorption images of an atom cloud in the MOT and in the transport trap illustrating the path of the atoms.} \label{fig:setup}
\end{figure}
The apparatus and the atom transport into the cavity are illustrated in Fig.\ref{fig:setup}. A cold cloud of $^{85}$Rb-atoms is prepared in a magneto-optical trap (MOT).
From there, these atoms are transferred over a distance of 14\,mm into a cavity by means of a horizontally running-wave dipole-force trap, which is formed by a single beam of an Yb:YAG laser.
This beam is red detuned from the relevant transitions 
and therefore attracts the atoms towards regions of highest intensity, i.e. towards its focus, which is located between MOT and cavity. The atoms oscillate with a period of 200\,ms in this trap,  
with one turning point at the position of the MOT, and the other in the cavity. This oscillation is stopped as soon as the atoms arrive in the cavity, by switching from the 
guiding dipole trap to a 
standing-wave dipole-trap 
(2 Watt, waist 16\,$\mu$m in the centre of the cavity).
The atoms are trapped in the antinodes of the standing wave that act as potential wells of 2.5\,mK depth, with a well-to-well distance of 515\,nm.
Nano positioning of the atoms relative to the cavity mode is accomplished by tilting a 
glass plate in the path of the retro-reflected beam. This changes the optical path length and therefore also the position of the antinodes. The glass plate is mounted on a galvo scanner that provides a 
measured repeatability of the interference pattern of  $\pm$\,15\,nm. This setup has been calibrated interferometrically. 
Since the trapped atoms follow this motion, they can be shifted to any position along the axis of the dipole-trap beams, as if they were sitting on a conveyor belt \cite{Kuhr01}. This allows a precise tuning of the coupling to the cavity. 

The cavity is $490\,\mu$m long so that transverse optical access is granted, and the mirror reflectivity is unbalanced, so that photons generated inside are emitted mainly through one of its mirrors. With the two mirrors (of 5\,cm radius of curvature) having transmission coefficients of $T_{0}=2\,$ppm and $T_{1}=95\,$ppm, strong coupling to the $TEM_{00}$-mode is reached with $(g_0, \kappa, \gamma)=2\pi\times(5, 5, 3)\,$MHz, and to the $TEM_{01}$-mode with $(g_0, \kappa, \gamma)=2\pi\times(4.3, 2.5, 3)\,$MHz. Here, g$_0$ is the atom-cavity coupling constant in a cavity antinode averaged over all the relevant atomic transitions, $\kappa$ the cavity-field decay rate and $\gamma$ the polarization decay rate of the atom. Note that we have deliberately chosen a regime (with $g_0\simeq\kappa$) where the photons are emitted from the cavity before they are reabsorbed due to vacuum-Rabi oscillations. Moreover, the cavity length (and therefore its mode volume) was chosen to be large enough to accommodate transverse pumping and trapping beams, resulting in $g_0\simeq\gamma$. We emphasize that single-photon generation schemes, like those demonstrated in Munich \cite{Kuhn02,Keller04}, work very well in such cavities. 

To continuously monitor the atom-cavity coupling, we tune the cavity resonance close to the $5^2S_{1/2}(F=3)$ to $5^2P_{3/2}(F'=4)$ transition and continuously excite the atom with a pair of counter-propagating lin$\perp$lin polarized laser beams. These pump laser beams run perpendicular to the cavity axis and have the same frequency as the cavity, so that resonant Raman scattering of photons into the cavity mode takes place, although the atoms are Stark-shifted out of resonance by the dipole trap. In particular, the transition 
\[
|F=3,n=0\rangle \stackrel{\mbox{laser}}{\longrightarrow} |F=4,n=0\rangle\stackrel{\mbox{cavity}}{\longrightarrow} |F=3,n=1\rangle
\]
keeps the atomic state unchanged, but changes the intra-cavity photon number, $n$, by one, so that the initial and final states differ, which is characteristic for Raman transitions. With the $TEM_{00}$-mode of the cavity driving one branch of this transition, the photon scattering rate from the atom-cavity system is in first approximation given by the enhanced spontaneous emission rate, $T_{1}\times c/(2L)\times(g_{\mbox{eff}}/\kappa)^2=640\,$ms$^{-1}$, with $g_{\mbox{eff}}=\Omega_l g_0/2\Delta_{ls}$ the effective atom-cavity coupling strength, where $\Omega_l=2\pi \times 30\,$MHz is the Rabi frequency of the driving laser and $\Delta_{ls}=2\pi\times 100\,$MHz the dynamic Stark shift from the trapping laser.

As we show elsewhere \cite{Nussmann05}, this cavity and pump beam configuration provides efficient cooling of the atoms in the centre of the cavity mode. Here, this is used to efficiently load atoms into the cavity. When the dilute cloud of atoms guided from the MOT arrives inside the cavity, the standing-wave potential is turned on. This gives rise to a random distribution of atoms among various potential wells in the vicinity of the cavity mode. Some atoms are hot enough to ripple along the dipole-trap axis like a marble across a washboard, in a random walk. As soon as such an atom enters the cavity mode, it scatters photons into the cavity and is subject to light forces which cool the atom into a potential well. The atom then remains trapped inside the cavity, where it continues to scatter photons. One after another, atoms are loaded into (or lost from) the cavity. This manifests itself in a stepwise increase (or decrease) of the photon-scattering rate into the cavity, which is therefore an unambiguous measure for the number of trapped atoms.

\begin{figure}
\centering\includegraphics[width=0.9\columnwidth]{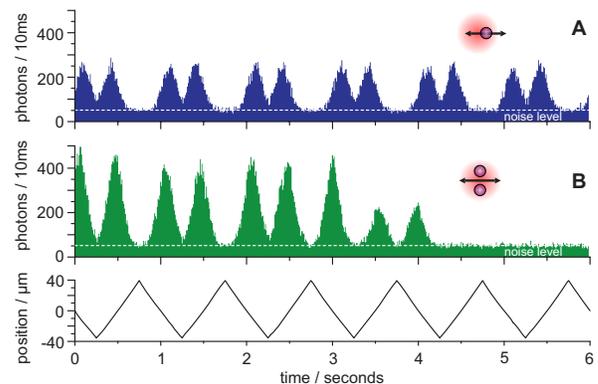}%
\caption{\textbf{Controlled coupling of individual atoms to the cavity:} Count rate of the photons emerging from the $TEM_{00}$-mode of the cavity while \textbf{(A)} a single atom is swept to and fro once a second over a distance of 75\,$\mu$m. \textbf{(B)} Same situation but starting with two atoms that are lost one after the other. }
\label{fig:Schieben00}
\end{figure}

To avoid a fluctuating atom number, we apply a filtering procedure. Some 10\,ms after the atoms have been loaded into the standing-wave trap, and when on average one atom has been caught, we interrupt the trap for 10\,ms and turn it back on adiabatically. During this interruption, only very cold and well-localized atoms remain captured in the weak dipole trap formed by the red-detuned light field that is used to stabilize the cavity length, while all other atoms are lost. The light of this additional trap is detuned from the Rubidium resonance by eight free spectral ranges of the cavity, i.e. by 5\,nm. Moreover, its 44\,$\mu$K-deep potential wells show a good overlap with the resonant cavity mode in the centre of the cavity. This helps to localize cold atoms in the antinodes of the cavity mode. Our preparation sequence 
usually results in 0, 1, or 2 atoms coupled to the cavity. From the photon scattering rate, we can determine the exact atom number within a few milliseconds.
Note that the weak intra-cavity trap together with the standing-wave trap forms a 2D optical lattice. The atoms finally trapped therein have a lifetime of 3\,s (without pump laser) and of more than 15\,s if continuously excited by the pump-laser. This is an unprecedented trapping time for a neutral atom under permanent observation inside a cavity.

As we load a well-known number of atoms into the cavity, we can also adjust the atom-cavity coupling to any value by tilting the glass plate. This is demonstrated in figure
\ref{fig:Schieben00}A, where a single atom is swept to and fro once a second over a distance of 75\,$\mu$m through the cavity mode. Figure \ref{fig:Schieben00}B shows the same situation, but starting with two atoms (twice the peak-count rate) that are lost one after another. This shows that we are able to manipulate a single atom for many seconds.

\begin{figure}
\centering\includegraphics[width=0.9\columnwidth]{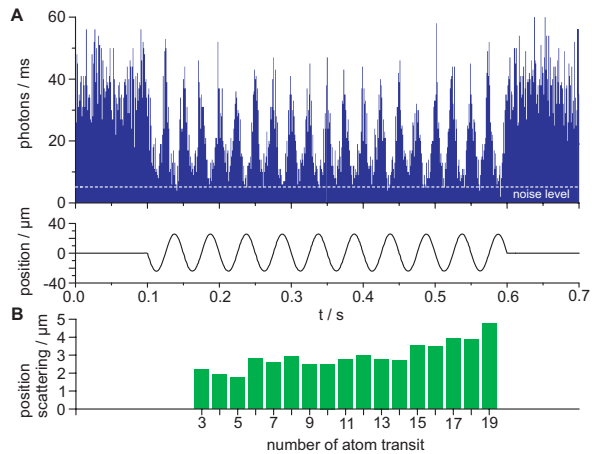}%
\caption{\textbf{Repeatability of atomic positions:} Count rate of photons emerging from the cavity while the atomic position is modulated with 20\,Hz over a distance of $\pm25\,\mu$m, and finally brought back to its initial value \textbf{(A)}. The 5\,kHz noise level is due to photons of the cavity dipole trap at 785\,nm still passing the 780\,nm filters used for shielding the photon counters.
\textbf{(B)} RMS deviation from the initial atomic position, averaged over 71 single-atom traces. The $\pm 2\,\mu$m initial spread stems from noise, but the gradual 135\,nm increase from transit to transit reflects the repeatability of the positioning scheme.}
\label{fig:posi}
\end{figure}

To demonstrate and analyze the repositioning capabilities of our setup, we have performed a series of experiments where we modulate the atomic position  with several  frequencies and amplitudes, and finally bring the atom back to its initial position. Figure \ref{fig:posi}A depicts the situation where the atom is initially sitting on the cavity axis. We sinusoidally sweep the atom to and fro for a period of 0.5\,s 
with a sweep amplitude of $25\,\mu$m. We stop this modulation at the initial position, and, as expected, the final photon count rate equals the initial count rate, i.e. the atom has been brought back to its starting point. We have now measured such traces for 71 individual atoms, and we swept each atom 19 times through the cavity. For each atom transit, we have determined the standing-wave position that leads to the optimum photon count rate. Figure \ref{fig:posi}B shows the mean deviation (RMS value) of these positions relative to the position found in the first or second transit (for both sweep directions separately). Due to the noise of our measurement technique, the  positions initially scatter over a range of $\pm 2\,\mu$m (in the third and fourth transit). However, the more important feature is the gradual increase of the mean deviation from transit to transit by an amount of 135\,nm. This slow increase is a measure of how well we can reproduce the atomic position. We therefore conclude that the atom is brought back to its starting point within $\pm 135\,$nm from one transit to the next. A possible reason for this uncertainty is the small but non-vanishing probability for the atom to hop to another potential well in the outer turning points. 

A similar method was applied to study the initial spatial distribution of the atoms, i.e. their average coupling to the cavity after the filter phase. A statistical investigation of 68 traces yields a width of the lateral distribution of $\pm$7.7\,$\mu$m along the axis of the dipole trap. This is small compared to the cavity waist, $w_0=29.5\,\mu$m. We therefore conclude that the average atom-cavity coupling is reduced by at most 7\,\% due to an initial lateral displacement. This indicates that during the filter phase the atoms had a temperature of only $5\,\mu$K in the $44\,\mu$K-deep cavity trap.

\begin{figure}
\centering\includegraphics[width=0.9\columnwidth]{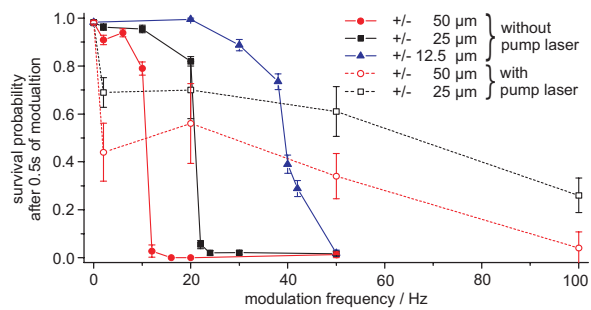}%
\caption{\textbf{Survival probability for moving atoms} after  0.5\,s
of sweeping to and fro, plotted as a function of the sweep frequency for three different sweep amplitudes. Measurements were performed with (open symbols) and without (closed symbols) pump laser. The different behavior reflects the influence of the pump laser on the dynamics of the atom.} \label{fig:Lifetime}
\end{figure}

Finally, we have investigated the effect of the atom transport on the lifetime with and without pump light. Figure \ref{fig:Lifetime} shows the probability for an atom to survive 0.5\,s of sweeping its position to and fro as a function of the sweep frequency for different sweep amplitudes. Without pump light, the survival probability is high for slow sweeps. However, a fast loss of atoms occurs as soon as the product of sweep frequency and amplitude exceeds $500\,\mu$m\,Hz. With the sweep being sinusoidal, this indicates strong losses if the maximum sweep velocity of the trap exceeds $2\pi\times 500\,\mu$m$/(\lambda_{YAG}/2)=6100$ potential wells per second. We know that residual back-reflections of the laser cause an intensity modulation at the same frequency. Therefore we attribute the sharp onset of losses to parametric heating, which is affecting the atom once the lowest trap frequency is met. 

If we carry out the same experiment with pump light, we see two effects: First, the survival probability is very high without modulation (0\,Hz), but it is reduced significantly even for low sweep frequencies, i.e. atoms are heated and can get lost once they leave the centre of the mode. Second, for high sweep frequencies, the strong cooling force in the centre of the cavity efficiently compensates for the parametric heating, as the atoms now have a significant chance to survive even for sweep frequencies around 100\,Hz. However, the data show that a loss-less controlled transport works best without pump light and with moderate velocity.

\begin{figure}
\centering\includegraphics[width=0.9\columnwidth]{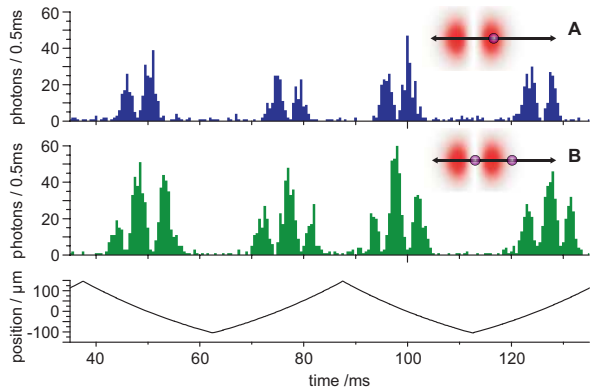} 
\caption{\textbf{Controlled coupling to the $TEM_{01}$-mode.}  (\textbf{A}) Photon count-rate from the $TEM_{01}$-mode while a single atom is swept to and fro with a frequency of 20\,Hz over a distance of 250\,$\mu$m. (\textbf{B}) Same situation, but starting with two atoms whose distance corresponds exactly to the distance between the two maxima of the $TEM_{01}$-mode. The probability for this arrangement is strongly enhanced due to our
loading scheme.} \label{fig:Schieben01}
\end{figure}

For the coupling of two (or more) atoms, it is advantageous to use higher transverse modes of the cavity. Figure \ref{fig:Schieben01}A shows the trace of a single atom that is repetitively swept across the $TEM_{01}$ mode of the cavity over a distance of 250\,$\mu$m, at a sweep frequency of 20\,Hz . The cavity mode shows two transverse intensity maxima, and we can clearly distinguish the coupling of the atom to these two maxima on its way through the
cavity. The most pronounced feature, however, is the vanishingly small photon-count rate when the atom is between these two maxima. This illustrates that we can control the position of an atom so precisely that it decouples  from the cavity once we move it between two maxima of a higher-order mode.

Most promising is the extension of this scheme to two atoms coupled to the same cavity mode, which is shown in figure \ref{fig:Schieben01}B. If two atoms are initially captured, our loading scheme ensures a probability of 50\% to have them in different maxima of the $TEM_{01}$ mode profile. It is now sufficient to sweep them to and fro along the dipole-trap axis to couple either each atom individually or both simultaneously to the cavity mode. The atoms can therefore be addressed individually via the cavity or, as the distance between the mode maxima of 42\,$\mu$m can be easily resolved with optical microscopes, 
one can address them individually from the side using independent laser pulses.

The degree of control achieved over the atom-cavity coupling is a large step towards quantum information processing with a quantum register consisting of neutral atomic qubits, where each individual atom can be addressed by and strongly coupled to a cavity \cite{Pellizzari95,Kuhn02,McKeever04,Keller04,Legero04}. In particular, a cavity like ours (with unbalanced mirror reflectivity and an adjustable number of atoms in it) is ideally suited to generate single photons from one atom \cite{Kuhn02,McKeever04,Keller04} or two photons from two atoms etc., with long coherence time  \cite{Legero04}. Moreover, the unitarity of the employed photon generation scheme is a way to deterministically entangle an atom and a photon, so that long-distance entanglement and teleportation between selected pairs of atoms in separate cavities are now possible in principle \cite{Cabrillo99,Bose99,Browne03}. We are confident that our positioning scheme also works in a cavity with much smaller losses. Here, the deterministically controlled coupling of two atoms to the same mode (as demonstrated above) is a sine qua non for cavity-mediated quantum-gate operations \cite{Pellizzari95,Rauschenbeutel99}.

\begin{acknowledgements}
This work was supported by the Deutsche Forschungsgemeinschaft (SPP 1078 and SFB 631) and the European Union (IST (QGATES) and IHP (CONQUEST) programs).
\end{acknowledgements}

\end{document}